\documentclass[twocolumn,tighten,times]{aastex62}
\pdfoutput=1

\submitjournal{ApJ}

\begin{document}
\title{Timing Properties of Shocked Accretion Flows around Neutron Stars - II. Viscous Disks and Boundary Layers}

\correspondingauthor{Ayan Bhattacharjee}
\email{ayan12@bose.res.in, sandipchakrabarti9@gmail.com}
\author[0000-0002-2878-4025]{Ayan Bhattacharjee,}
\affiliation{S. N. Bose National Centre for Basic Sciences, Salt Lake, Kolkata 700106, India}
\author[0000-0002-0193-1136]{Sandip K. Chakrabarti}
\affiliation{Indian Center for Space Physics, 43 Chalantika, Garia St. Road, Kolkata 700084, India}

\label{firstpage}
\begin{abstract}
We use Smoothed Particle Hydrodynamics to study viscous accretion flows around a weakly magnetic neutron star. We show the formation of multiple ``boundary" layers in presence of both cooling and viscosity. We find that with the introduction of a small viscosity in a sub-Keplerian flow, much like the wind accretion in HMXBs such as Cir X-1, only a single Normal Boundary Layer (NBOL) forms to adjust the rotational velocity component. With the increase of viscosity, the region extends radially and beyond some critical value, a RAdiative KEplerian Disk/layer (RAKED) forms between the sub-Keplerian flow and the NBOL. When viscosity is increased further only NBOL and RAKED remain. In all such cases, the CENtrifugal pressure dominated BOundary Layer (CENBOL) is formed, away from the star, as in the case of black holes. This is the first self-consistent study where such a transition from sub-Keplerian flows has been reported for neutron stars. We also identify the connection between accretion and ejection of matter, following the Two-Component Advective Flow for black holes, for neutron stars. The results are crucial in the understanding of the formation of disks, boundary layers and outflows in wind dominated neutron star systems.
\end{abstract}


\section{Introduction}
In Bhattacharjee \& Chakrabarti 2019 (hereafter Paper I), the study of accretion flows around weakly magnetic neutron stars is divided into two broad classes: 1. The flow is inviscid, advective and has a low efficiency of radiation. It also comprises of winds from the companion star. 2. The flow has a significant viscosity to redistribute the specific angular momenta. The authors discussed the Class 1 flows in Paper I, i.e., the effects of angular momentum of a sub-Keplerian flow in absence of significant dynamic viscosity. The less-explored domain of sub-Eddington, inviscid accretion around neutron stars (NSs) was studied in detail using hydrodynamic simulations. Paper I also argues why a Two-Component Advective Flow or TCAF (Chakrabarti 1995, 1997) solution is most likely the generalized solution for accreting matter around a weakly magnetic neutron star as well. For a detailed review of the studies, refer to Bhattacharjee 2018, and the references of Paper I. The TCAF solution has been used to explain the spectral and timing properties of stellar mass black holes in a self-consistent manner for multiple transient sources (Debnath et al. 2014, Bhattacharjee et al. 2017, Shang et al. 2019 and the references therein), for persistent sources such as Cygnus X-1 (Banerjee et al. 2019), for class variable sources such as GRS1915+105 (Banerjee et al. 2019). The paradigm can also be applied to NSs with some modifications (Bhattacharjee \& Chakrabarti 2017, hereafter BC17). 

For inviscid flows with sub-Keplerian angular momentum, two axi-symmetric shocks are present in an accretion flow. The outer shock is the so-called 
CENtrifugal pressure dominated BOundary Layer or CENBOL (Chakrabarti 1995, 1997) and the inner one is the NBOL (BC17, Chakrabarti 2017, Bhattacharjee 2018, Paper I). We note below some of the key findings reported in Paper I.

\textbf{Origin of Quasi Periodic Oscillations (QPOs):} The simulations produce both low and high-frequency QPOs and the oscillations last during the whole simulation period (more than 200 dynamical timescales measured at the injected flow radius, i.e., $30~r_s$). This suggests that the QPOs are formed due to a part of the flow dynamics and not a transient effect as inferred by others (e.g., Belloni et al. 2002; Mauche 2002; Barret and Olive, 2005; Mendez 2006). Both the centroid frequencies and Q factors match well with observed results of neutron stars such as GX17+2, 4U 1728-34 and Cir X-1. It implied that the advective flow suggested in the literature while explaining the behavior of the source Cir X-1 (Boutloukos et al. 2006), may be the same as the dynamic transonic flow solution we discuss here. Paper I showed that the presence of angular momentum itself can generate multiple modes of oscillation in CENBOL and NBOL, manifesting as QPOs in the PDS, in presence of cooling.
In fact, the ratio of different high frequency QPOs, the separation between the peaks and the Q factors of different QPOs were also well in agreement with the observed cases (Barret et al. 2005, Boutloukos et al. 2006). 

\textbf{Spin-up rate of NS:} The spin-up torque $(N)$ measured from the simulated results of Paper I was such that it corresponded to a spin-up rate of $\sim 5.0 \times 10^{-14} Hz~s^{-1}$. The value agrees well with observational results and predictions from other models (Bildsten 1998; Revnivtsev \& Mereghetti, 2015; Sanna et al. 2017; Bhattacharyya \& Chakrabarty, 2017; G{\"u}gercino{\v g}lu \& Alpar, 2017; Ertan 2018).

\textbf{Accretion-Ejection:} The solution captured the connection between accretion and outflows in the sense that 
the  CENBOL and NBOL launched outflows in both the quadrants. The interaction of inflow and outflow lead to the formation of bending instabilities (reported in M01a for black holes) and correspond to the hecto-Hz oscillations found in the PDS. It was also shown that a decrement of angular momenta (from $1.8$ to $1.7$) reduced the oscillating nature as shown by Deb et al. (2016) for black holes.

In the present paper, we focus on the effects of the introduction of viscosity and its gradual enhancement on the flow configuration from a theoretical point-of-view. We note the similarities and differences with the cases of Paper I. We study the formation of disks, boundary layers in the inflow. We also note the changes in the pattern of ejection of matter from the equatorial plane as compared to that in inviscid flows studied in Paper I.

\begin{figure*}
\centering
\includegraphics[height=6.0cm,width=4.25cm]{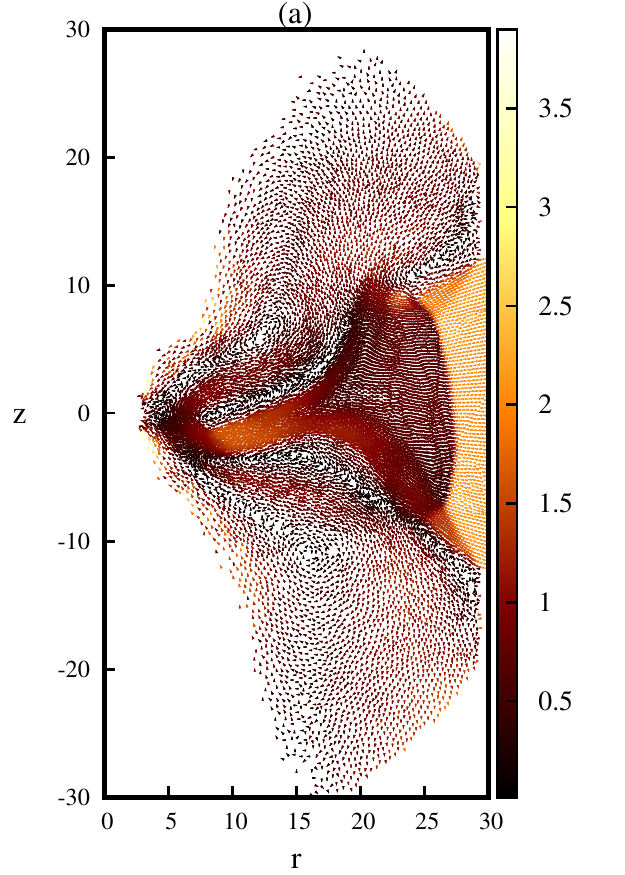}
\includegraphics[height=6.0cm,width=4.25cm]{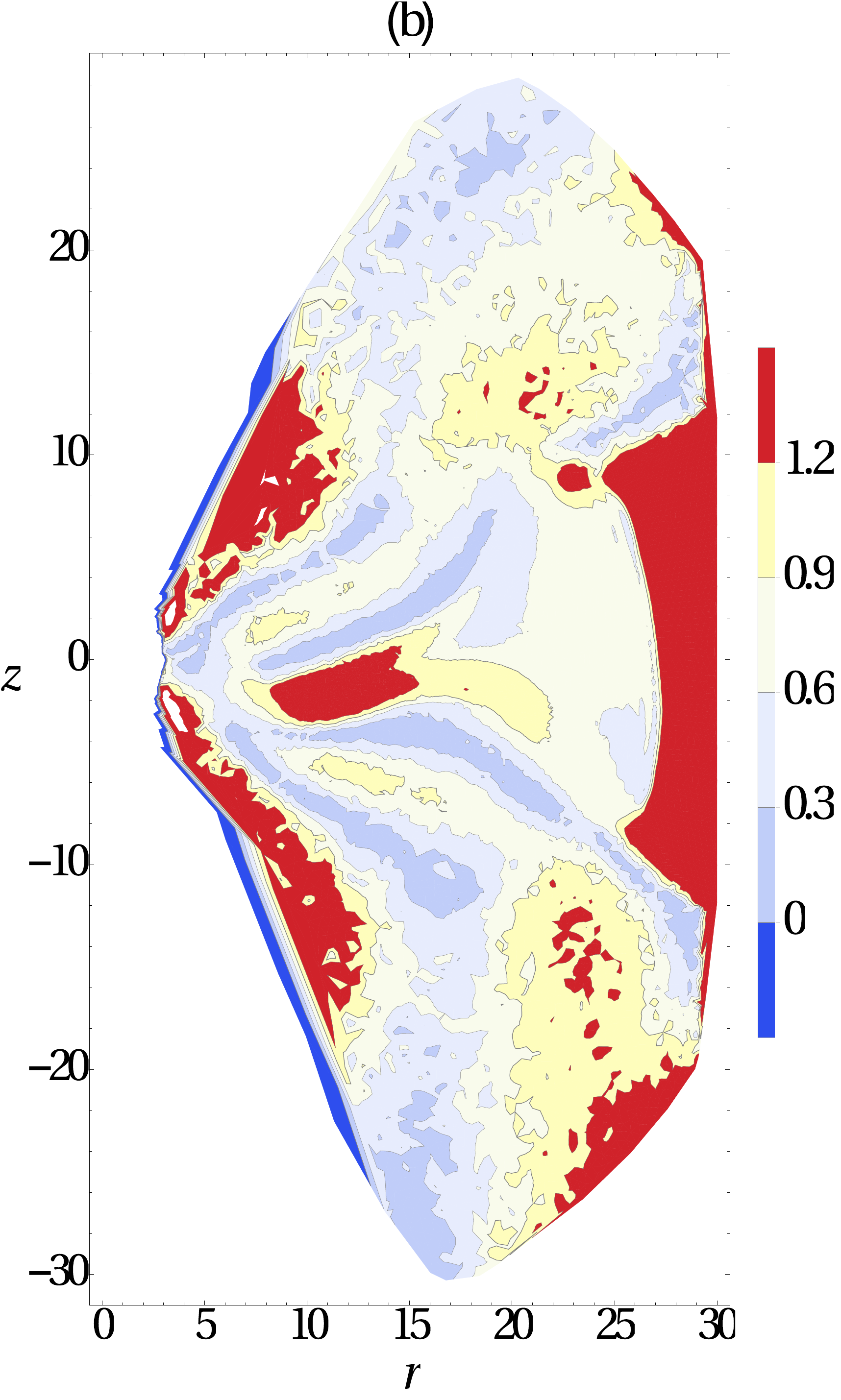}
\includegraphics[height=6.0cm,width=4.25cm]{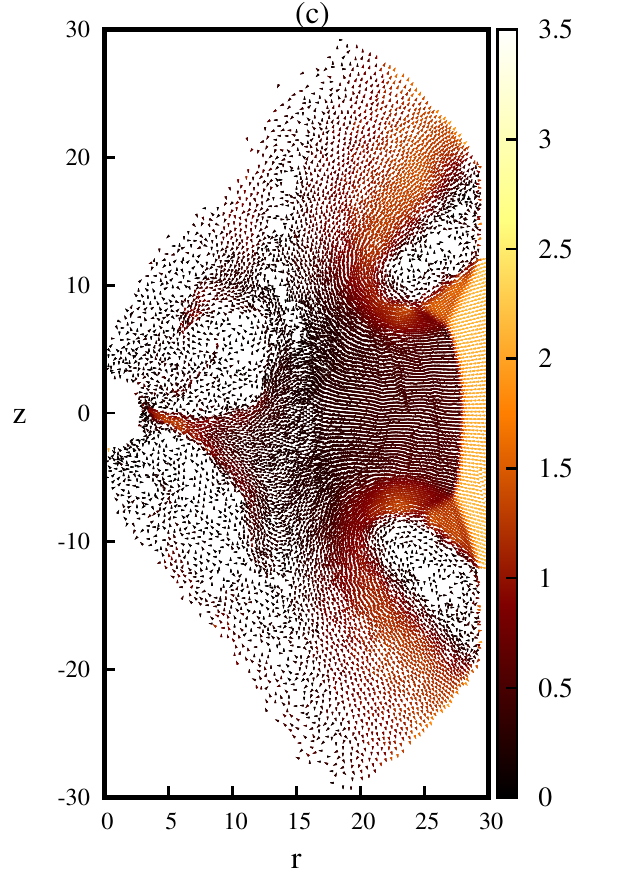}
\includegraphics[height=6.0cm,width=4.25cm]{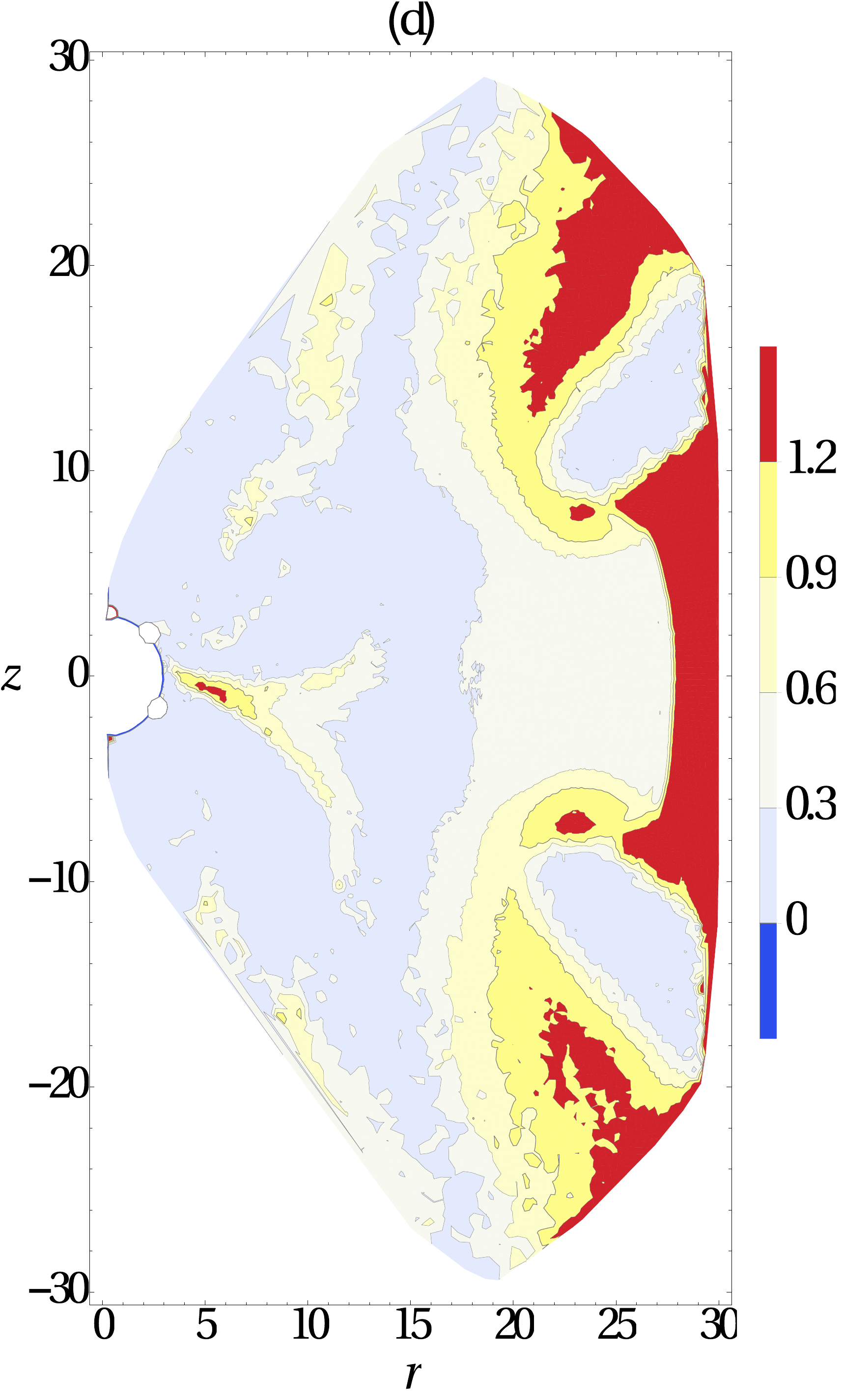}
\includegraphics[height=6.0cm,width=4.25cm]{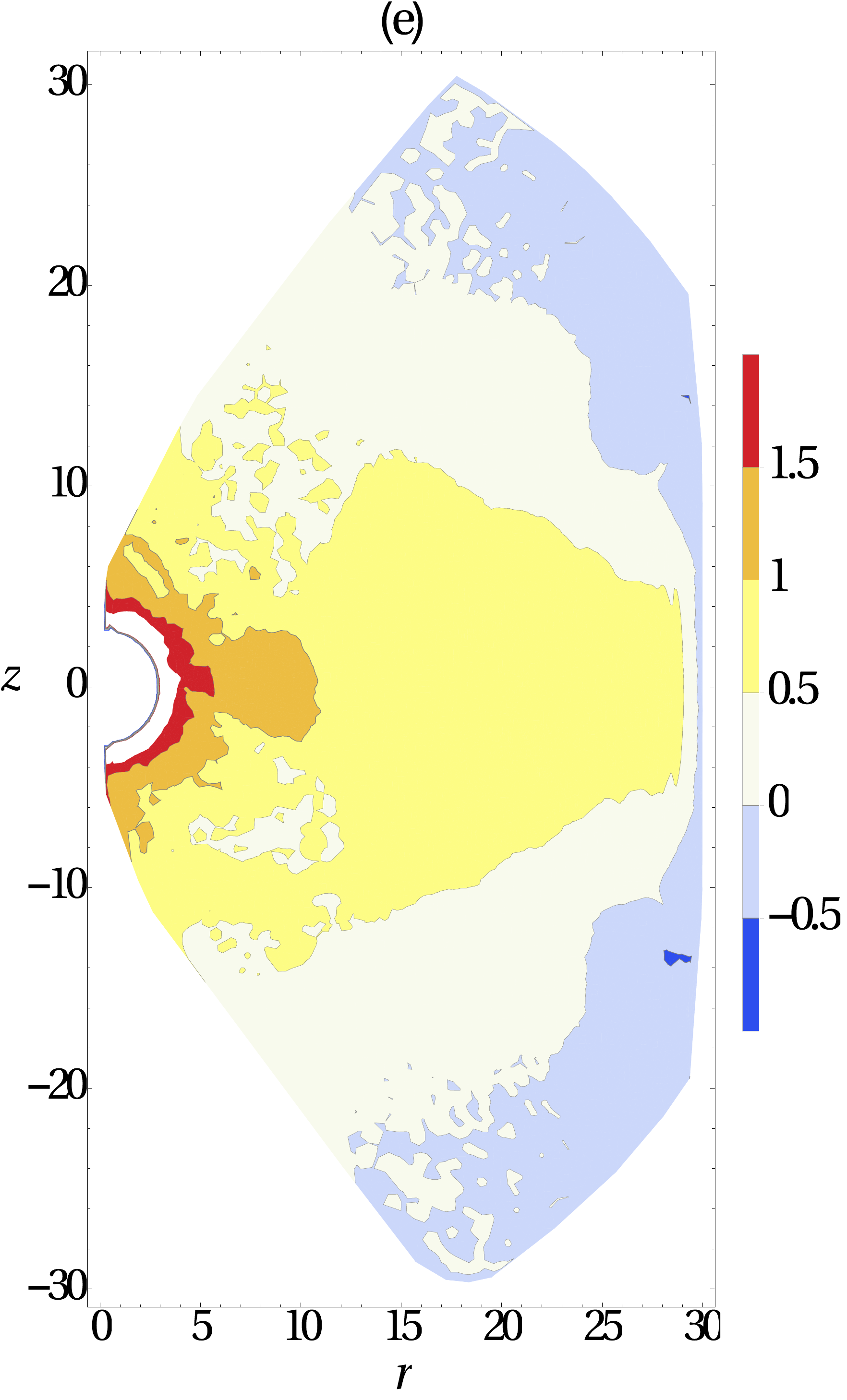}
\includegraphics[height=6.0cm,width=4.25cm]{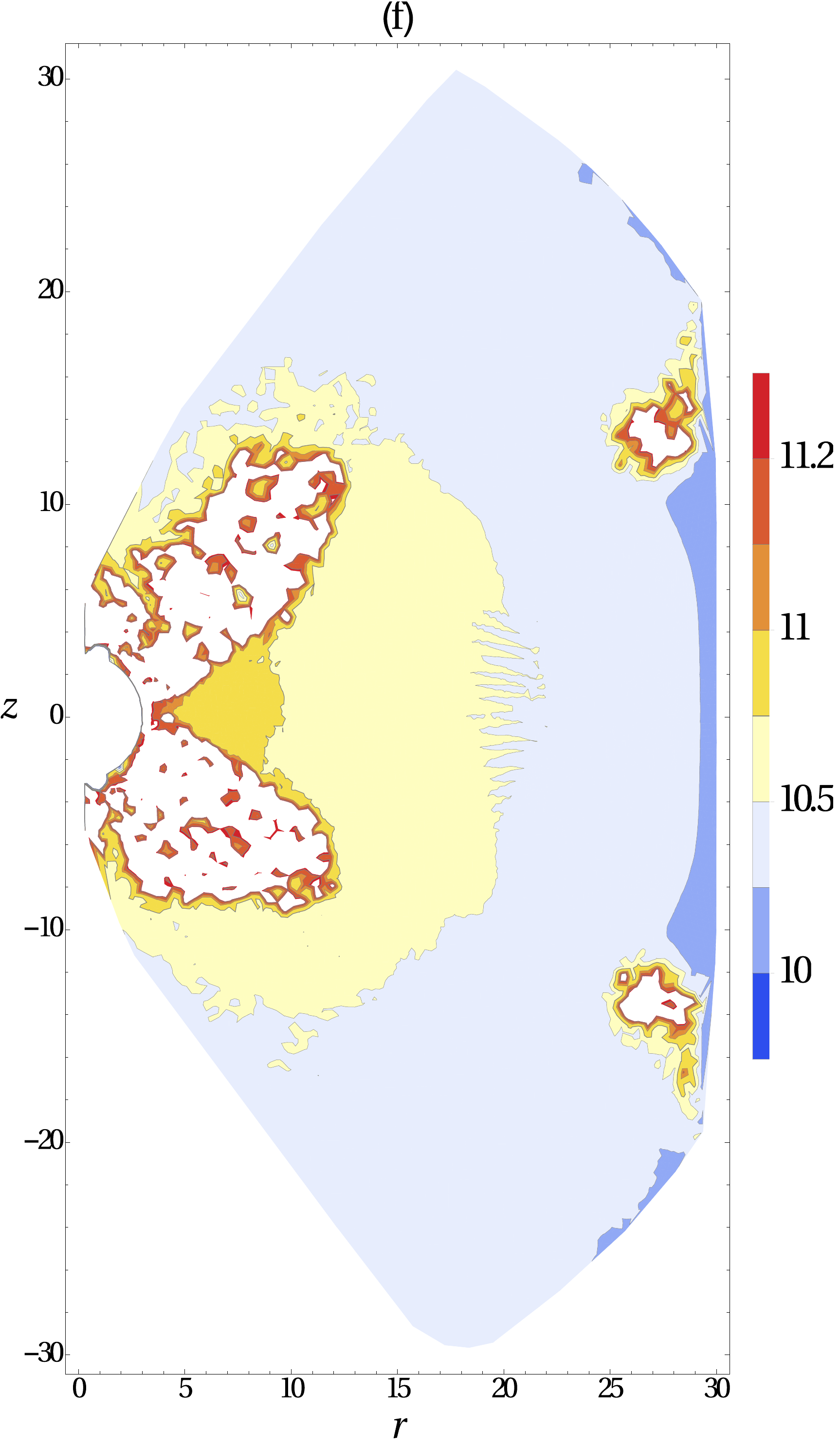}
\includegraphics[height=6.0cm,width=4.25cm]{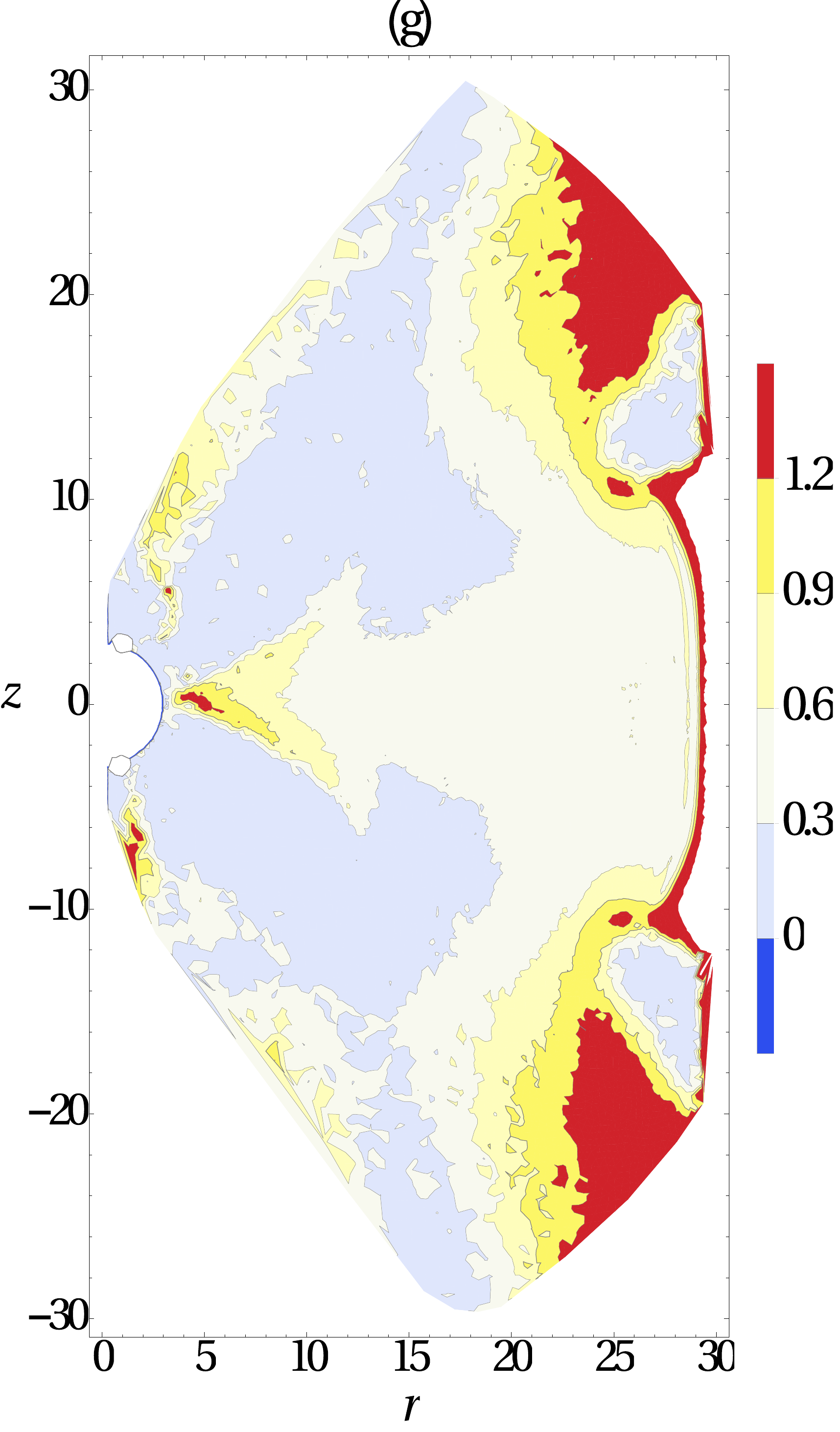}
\includegraphics[height=6.0cm,width=4.25cm]{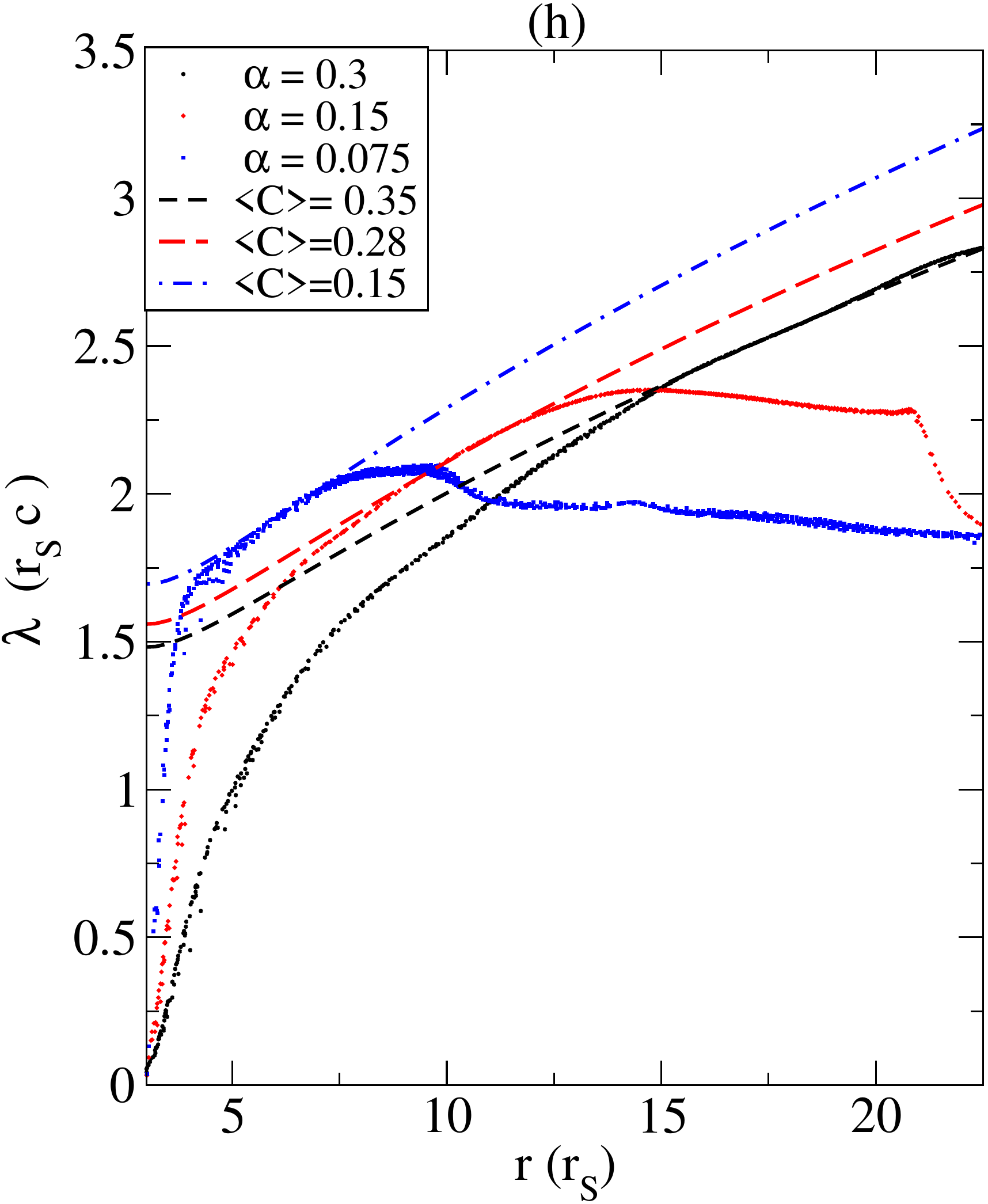}
\caption{(a) Velocity vector $v_r \hat{r} + v_z \hat{z}$ (arrow heads) with Mach number in the colour bar, for the flow configuration $C1$ at time $t=0.2800 s$ and (b) corresponding contours of constant Mach number range showing the outer and inner shocks. (c) Velocity vector $v_r \hat{r} + v_z \hat{z}$ (arrow heads) with  Mach number in the colour bar, for the flow configuration $C2$ at time $t=0.1400 s$ and (d) corresponding contours of constant Mach number range showing the outer and inner shocks. (e) Contours of constant $log(\rho/\rho_0)$ range for case $C3$ at time $t=0.0400 s$ and (f) Corresponding contours of constant temperature range in $K$ (log scale) and (g) Corresponding Mach number range contours. (h) Formation of different regions in inflow for various $\alpha$, at a time $t=0.016 s$. Dashed curves are the theoretical distribution with $\frac{\sqrt{1-<C>}r^{1/2}}{\sqrt{2}(1-1/r)}$ shape, while the dotted curves are as obtained from the simulations. Clear indication of RAKED formation is seen.} 
\label{fig1}
\end{figure*}

\section{Method}
The Smoothed Particle Hydrodynamics (SPH) method was introduced by Monaghan (1992). It has since been used in many astrophysical systems, including simulation of accreting matter around black holes to simulate accretion in 1D (Chakrabarti \& Molteni 1993, hereafter CM93); 2D flows (Molteni, Lanzafame \& Chakrabarti 1994, hereafter MLC94); viscous Keplerian discs (Chakrabarti \& Molteni 1995, hereafter CM95); resonance oscillation of shocks due to cooling in 2D (MSC96); comparative study of shocked advective flows using SPH and TVD schemes (Molteni, Ryu \& Chakrabarti 1996, hereafter MRC96); thick accretion discs (Lanzafame, Molteni \& Chakrabarti 1998, hereafter LMC98); bending instability of an accretion disc (Molteni et al. 2001a, hereafter M01a; Molteni et al. 2001b, hereafter M01b); interaction of accretion shocks with winds (Acharyya, Chakrabarti \& Molteni 2002, hereafter ACM02); and effect of cooling on the time dependent behavior of accretion flows (Chakrabarti, Acharyya \& Molteni 2004, hereafter CAM04).

We use the SPH code used in Paper I to solve for the conservation of mass, momentum and energy. All the equations, notations and boundary conditions (inner and outer) are kept the same. We add the $\alpha$-viscosity prescription for the system following LMC98, which modifies the following.

1. Conservation of azimuthal component of the momentum:
\begin{equation}
\footnotesize{\Big{(} \frac{D v_{\phi}}{Dt} \Big{)} = - \Big{(} \frac{v_{\phi}v_r}{r} \Big{)} + \frac{1}{\rho}\Big{[} \frac{1}{r^2} \frac{\partial}{\partial r} (r^2 \tau_{r\phi}) \Big{]},}
\end{equation}
where, $\tau_{r\phi}=\mu r \frac{\partial \Omega}{\partial r}$, $\Omega=\frac{v_{\phi}}{r}$. Here, we use the standard Shakura-Sunyaev turbulent viscosity (Shakura \& Sunyaev 1973) $\mu=\alpha \rho a Z_{disk}$, where $Z_{disk}$ is the vertical thickness of the flow as obtained from the vertical equilibrium condition and is given by $Z^2_{disk}=\frac{2}{\gamma} a r (r-1)^2$ (Chakrabarti 1989, 1990; LMC98).

2. Conservation of energy (viscous heating term added)
\begin{equation}
\frac{D}{Dt}\Big{(}e+\frac{1}{2}\vec{v}^2\Big{)}=-\frac{P}{\rho} \vec{\nabla} \cdot \vec{v} + \vec{v} \cdot \Big{(}\frac{D\vec{v}}{Dt}\Big{)} - \zeta_{1/2} \rho e^{\alpha} + \frac{\mu}{\rho}\Big{[}r \frac{\partial \Omega}{\partial r}  \Big{]}^2
\end{equation}

Equations 1 and 2 are identical to the Eq. 15 and 6 of Paper I, apart from the last terms which introduce the viscous effects in the flow (in Paper I, $\mu$ was 0). The flow was injected from $r_{inj}=30r_S$ with $v=0.1211$, $a=0.0590$ and $\lambda_{inj}=1.7$ (similar to case C2 of Paper I).

\section{Results}
We increased the viscosity parameter from $0.075$ (C1) to $0.15$ (C2) to $0.3$ (C3) and kept injected $\lambda_{inj}=1.7$ (see, Table 1). We also define the RAdiative KEplerian Disk as the equivalent of a standard Keplerian Disk following the prescription of Chakrabarti \& Sahu 1997 (hereafter, CS97), when the effect of the radiative pressure term due to the emission from NBOL is included. If the average repulsive radiative force is $F_{rad}=\frac{<C>}{2(R-1)^2}$, then the effective gravitational force reduces to $F_{g}=\frac{1-<C>}{2(R-1)^2}$ (CS97, Paper I). 

\textbf{Case C1:} In presence of lower viscosity ($\alpha=0.075$), the sub-Keplerian flow adjusted its angular momentum very close to $R_{NS}$. The layer where most of this transition took place was the previously identified Normal Boundary Layer (NBOL). In a very small region ($5-7~r_S$), a RAdiative KEplerian Disk (RAKED) appears and disappears from time to time. Figure 1(h) shows one such instance when the RAKED is formed. Rest of the outer flow remained sub-Keplerian in nature. This is similar to the cases reported in Paper I. We also see that the turbulent nature of the inner flow, which shows multiple bending instabilities, is very similar to the cases reported in Paper I. Furthermore, matter is ejected from both CENBOL and NBOL in the upper quadrant and only from CENBOL in the lower quadrant. The outflowing matter from NBOL undergoes multiple shock transitions (one before merging with CENBOL-outflow, one after), before becoming transonic near the outer edge. A part of the ejected matter falls back to CENBOL and another part is accreted onto the NS through more radial shocks. This fallback onto the star is primarily achieved due to the lowered angular momentum near the surface.

\textbf{Case C2:} When the viscosity parameter is increased from $0.075$ to $0.15$, the size of the NBOL is increased from $\sim 2 r_S$ to $\sim 5.5 r_S$. The size of RAKED is also increased 
from $\sim 2.5 r_S$ in C1 to $\sim 3.5 r_S$ here, though it has relatively weaker oscillations. The outer part of the flow remained sub-Keplerian throughout the simulation. This case is of particular interest as all the layers are present simultaneously. When the velocity vectors and Mach numbers are compared with those in C1, we notice that the flow has become more stable due to introduction of viscosity and consequent decrease of specific angular momentum in the inner regions. The outer shock is moved at a larger distance from the star boundary. The NBOL did not eject any matter due to the lack of centrifugal drive
by low angular momentum and only the CENBOL ejected matter. A part of the outflowing matter is seen to move towards the NS surface transonically. The flow interacts with a larger area on the NS surface due to the further reduction of angular momentum near the star. 
The inner shock location moves closer to the NS and is narrower.

\textbf{Case C3:} Further increase of viscosity parameter from 0.15 to 0.3 leads to increase of NBOL size to $\sim 11.5 r_S$. Furthermore, the RAKED kept growing in size as its outer boundary started to increase with increasing time. The inner edge of RAKED (outer edge of NBOL) remained fairly constant. We also show the density, temperature and Mach number contours of this case at time $t=0.04 s$, in Figs. 1(e), 1(f) and 1(g), respectively. Both the density jumps near CENBOL and NBOL are distinctly seen in 1(e) and 1(g). We also notice that matter is ejected from near the the outer shock. The temperature distribution of the flow also captures the shock transition. In addition to those, further hotter and clumpy regions are seen near the NBOL where the inflow and outflows mix. In this case, the shock is pushed to the outer boundary very quickly and matter is ejected in both quadrants from CENBOL. NBOL is almost symmetric about $z=0$ and did not produce any outflow. Fallback of some matter onto the star is also seen. 

\begin{table*}
\centering
\vspace{0.5cm}
\normalsize{
\begin{tabular}{c c c c c c c}
\hline
\hline
Case & $\alpha$ & $<C>$ & NBOL ($r_S$) & RAKED ($r_S$) & SK ($r_S$) & $R_{CE,O}$ ($r_S$)\\
\hline
\hline
C1 & 0.075 & 0.15 & $R_{NS}-5.0$ & $5.0-7.5$ & $7.5-R_{out}$ & $17.5$\\
\hline
C2 & 0.15 & 0.28 & $R_{NS}-8.5$ & $8.5-12.0$ & $12.0-R_{out}$ & $21.0$\\
\hline
C3 & 0.3 & 0.35 & $R_{NS}-14.5$ & $14.5-23.5$ & - & $27.5$ \\
\hline
\hline
\end{tabular}
}
\caption{Different regions of the inflow for different values of $\alpha$, at $t=0.016~s$, for $-0.5<z<0.5$. Here, $R_{CE,O}$ is the outer boundary of CENBOL or the outer shock location.}
\end{table*}

The $\lambda$ vs $r$ curve is drawn in Fig. 1(h) to show the region where Raked is formed. The dashed curves are theoretical shapes $\lambda = \frac{\sqrt{1-<C>}r^{1/2}}{\sqrt{2}(1-1/r)}$ while the dotted curves are as obtained from simulations. The values of $<C>$ increased with the increase of $\alpha$, suggesting that more energetic matter (due to viscous heating) makes its way to the NS surface and raised the radiative pressure exerted on the inflowing matter. 

\section{Conclusions}
In this paper, we report the formation of a normal boundary layer and a radiative Keplerian disk out of a sub-Keplerian injected flow around a weakly magnetic neutron star. To our knowledge, this is the first detailed study of viscous flows around a weakly magnetized neutron star. We carried out a qualitative study of the behaviour of three regions formed in the inflowing matter as the flow viscosity is gradually increased. When viscosity is low, resembling an inviscid flow, matter adjusts its angular momenta only very close to the NS surface and forms the NBOL out of the sub-Keplerian flow. RAKED is transient in nature for these cases. We believe this happens in HMXBs, such as, Cir X-1. We notice that two regions, one on the inner edge of CENBOL and the other near NBOL, eject matter along the z-direction. A part of these two matter merge and a part actually falls back to these regions. For higher viscosity, the RAKED is formed in between NBOL and the sub-Keplerian flow. Here, the RAKED is small and steadily oscillate. These would be the most general type of inflow in presence of viscosity. We also notice that the redistribution of angular momentum helps in ejecting more matter out of the equatorial plane, even more compared to inviscid cases. For even higher viscosity, a clear RAKED is formed and increased in size towards larger radial distance. This suggests that these cases are possible in the super-critical range of viscosity. Within 25 dynamical timescales at $30 r_S$, the RAKED reached the outer edge. The redistributed angular momentum also leads to an even higher ejection of matter from the disk. The added viscosity appears to make the flow more stable and the vertical oscillations become negligible.

In Paper I, the simulations were carried out for more than 200 dynamical timescales measured at the injected flow radius, i.e., $30~r_S$, which was around $0.33 s$. In this paper, we carried out $C1$ and $C2$ till the same timescales as the solutions appear to be free from
boundary effects during this time. The flow configurations plotted in Figs. 1(a-d) are taken from that duration to showcase some typical features. For the case C3, the outer shock reached the outer boundary soon after $0.04 s$, and we only comment on the solutions prior to that time to avoid effects which could be artefacts of simulations. The flow configuration we discuss is at $t=0.04 s$. The $\lambda$ vs $r$ plot was made at a much shorter time ($0.016s$) to capture the effects of viscosity away from the outer boundary.

We also chose the simulation box to be the same as that of Paper I for a proper comparison. The system can be studied for a larger radial distances, which can be used to determine the critical viscosity $\alpha_c$ where the entire inflow, apart from NBOL, becomes a RAKED out of a sub-Keplerian matter flowing in. The study of the system on a larger vertical scales will enable us in tracking the ejected matter better. Whether the outflowing matter falls back onto the inflow, escapes, or catches up with a previous shock, can also be investigated by such a study on larger length-scales. Both of these cases would require the system to be studied for a much larger timescale and would require a robust computational setup and longer runtime. These are beyond the scope of the present paper and are being pursued at present to be reported in some future work. 
\section*{Acknowledgement}
AB acknowledges the computational support provided by SNBNCBS. SKC acknowledges support from the DST/SERB sponsored Extra Mural Research project (EMR/2016/003918) fund.

\end{document}